\documentclass[aps,prl,reprint,amsmath,amssymb,showpacs,superscriptaddress]{revtex4-1}
\usepackage[english]{babel}
\usepackage{amsmath,amssymb,graphicx,xcolor}
\usepackage{natbib}
\newcommand{\nobracket}{}
\newcommand{\tmcolor}[2]{{\color{#1}{#2}}}
\newcommand{\tmem}[1]{{\em #1\/}}
\newcommand{\tmop}[1]{\ensuremath{\operatorname{#1}}}
\newcommand{\tmtextit}[1]{{\itshape{#1}}}

\begin{document}

\title{Design of a Cooper pair box electrometer \texorpdfstring{\\}{}
for application to solid-state and astroparticle physics}

\author{L. Tosi}
\affiliation{CSNSM, Univ. Paris-Sud, CNRS/IN2P3, Universit{\'e} Paris-Saclay,
91405 Orsay, France}

\author{D. Vion}
\affiliation{SPEC, CEA, CNRS, Universit{\'e} Paris-Saclay, CEA Saclay, 91191
Gif-sur-Yvette, France}

\author{H. le Sueur}
\email[Corresponding author: ]{helene.le-sueur@csnsm.in2p3.fr}
\affiliation{CSNSM, Univ. Paris-Sud, CNRS/IN2P3, Universit{\'e} Paris-Saclay,
91405 Orsay, France}

\begin{abstract}
  We describe the design and principle of operation of a fast and sensitive
  electrometer operated at millikelvin temperatures, which aims at replacing
  conventional semiconducting charge amplifiers in experiments needing low
  back-action or high sensitivity. The electrometer consists of a Cooper Pair
  box (CPB) coupled to a microwave resonator, which converts charge variations
  to resonance frequency shifts. We analyze the dependence of the sensitivity
  on the various parameters of the device, and derive their optimization. By
  exploiting the nonlinearities of this electrometer, and using conventional
  nanofabrication and measurement techniques, a charge sensitivity of a few
  $10^{- 7} e / \sqrt{\tmop{Hz}}$ can be achieved which outperforms existing
  single charge electrometers. 
\end{abstract}

\maketitle

\section{Introduction}
The detection of the time variations of minute amounts of charge is
instrumental in the success of recent experiments in many fields. We can cite,
for instance, the measurement of the transport statistics in a mesoscopic
circuit {\cite{field_measurements_1993}}, the demonstration of exotic states
of matter such as Majorana fermions {\cite{ben-shach_detecting_2015}}, or even
the observation of new phenomena in astrophysics such as neutrino coherent
scattering {\cite{anderson_coherent_2011,akimov_observation_2017}} or the
search for direct dark matter detection at low masses
{\cite{edelweiss_collaboration_optimizing_2017}}.

Such ultimate charge sensing requires to operate at low enough temperatures in
order to suppress the thermal charge fluctuations that would mask the desired
phenomena. These experiments are typically performed in dilution refrigerators
at temperatures around 10-50 mK. In this temperature range, conventional
semiconducting charge amplifiers generally do no work or they dissipate more
than the refrigerator can handle. Secluding the readout electronics at higher
temperature is neither desired because it unavoidably adds stray capacitances
that reduce the charge-detection bandwith and increase the noise pick-up.

It has long been shown that single charge mesoscopic devices (SCD) are well
suited for ultimate electrometry at these temperatures. Some of these devices
are actually the dual of the SQUID and as such, they are also potentially
quantum limited. The most famous SCDs are the Single Electron
Transistor or quantum dots that are operated essentially as field transistors:
the gate charge modulates the out-of-equilibrium (dissipative) current flowing
between source and drain. Such out-of equilibrium operation creates many
electron-hole pairs nearby the SCD, that will eventually relax by emitting
phonons or photons into the environment. These radiations in turn influence
the charge to be detected in a sensitive device. It is often difficult to
assess the influence that those microscopic excitations can have onto the
measured systems. Fortunately, such uncontrolled electromagnetic back-action
can be completely avoided by using superconducting versions of SCDs, such as
the single Cooper pair transistor (CPT) or its simpler version the Cooper pair
box (CPB). Indeed, these devices can be operated in equilibrium (no voltage
bias and no dissipation), making their back-action minimal and through a
well-identified channel.

Finally, large-bandwidth implementations {\cite{schoelkopf_radio-frequency_1998,devoret_amplifying_2000}} of these
mesoscopic charge sensing devices were demonstrated by coupling them with
resonant microwave ($\mu$w) circuits. This further enables the
frequency-multiplexing of charge detectors on a single line which is highly
desirable in experiments where many charge channels need to be monitored.
Hence our choice to use an optimized microwave superconducting, i.e. non
dissipative, electrometer, which has both large bandwidths and multiplexing
capabilities, and with a minimal back-action.

In this work we analyse in depth the optimal design of a $\mu$w Cooper
pair box electrometer suitable for a large variety of applications, including
the charge channel of bolometric particle detectors. Our optimization takes
advantage of the quantum nature of the coupled resonator - CPB system and it
includes the analysis of non-linearities, that are shown to enhance the
performances by one order of magnitude compared to state-of-the-art
{\cite{brenning_ultrasensitive_2006,naaman_time-domain_2006,de_graaf_charge_2013}}.

In section 2 we describe our electrometer based on a CPB embedded in a
microwave resonator. The electrometer's small signals sensitivity is derived
in section 3, first in the linear regime of the CPB, then taking into account
the CPB nonlinear response, and finally including the resonator quantum
nonlinearities. Section 4 addresses the main sources of imperfections, i.e.
the problem of offset charge noise and quasiparticle poisonning of the island
{\cite{lafarge_measurement_1993}}. Section 5 is a brief discussion about how
to integrate our electrometer in a detector.

\section{The RF-CPB electrometer}

\subsection{The Cooper pair box}

\begin{figure}[!t]
  \resizebox{1\columnwidth}{!}{\includegraphics{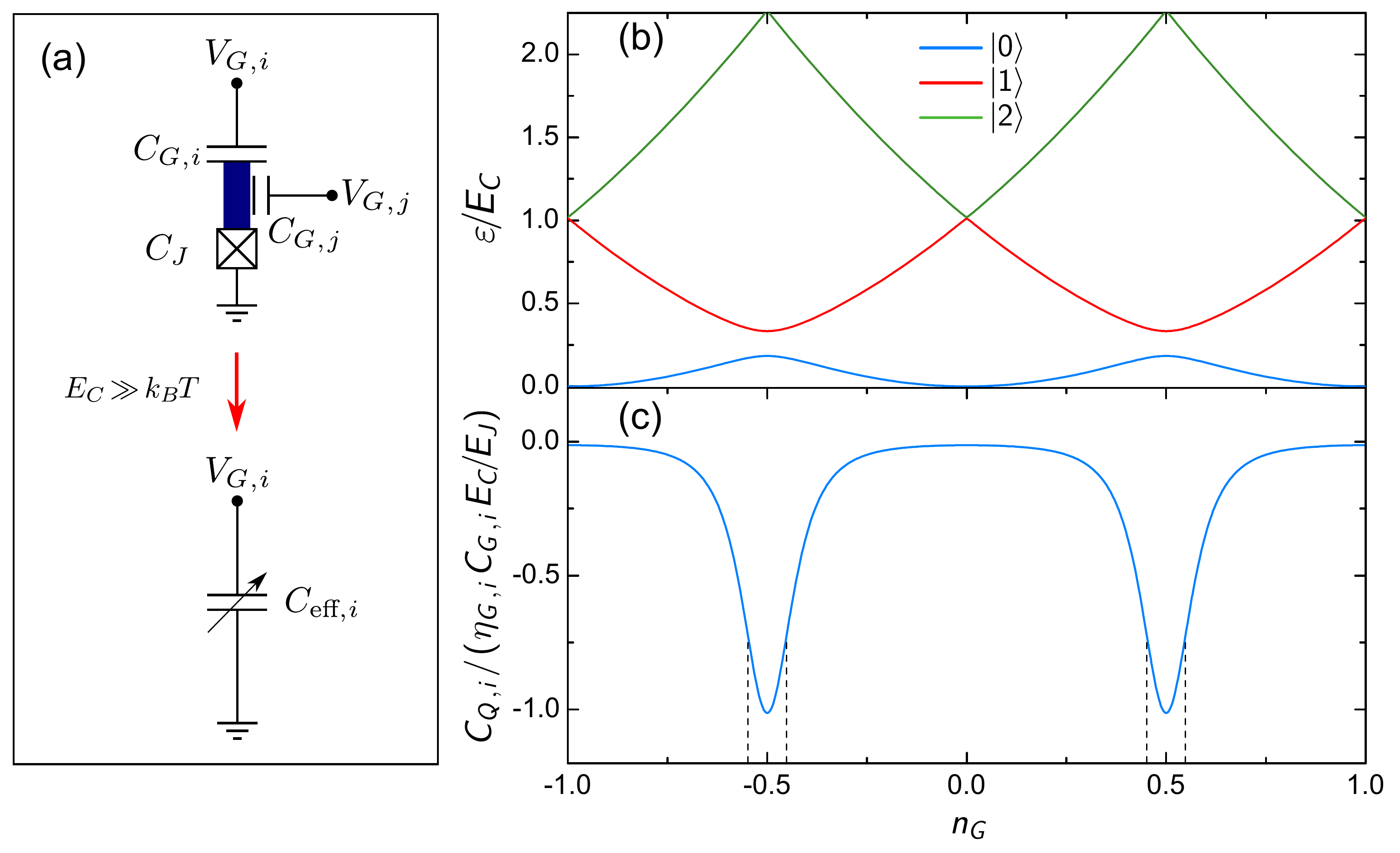}}
  \caption{\label{NRJ_Cq_of_nG}The CPB and its effective capacitance. (a)
  Circuit representation of the CPB: an island (dark rectangle) connected to
  the ground through a Josephson junction, and capacitively coupled to one or
  several ``gate'' electrodes. As seen from a particular gate $i$ the circuit
  is equivalent to an effective variable capacitance $C_{\tmop{eff}, i}$ when
  $E_C$ is large compared to $k_B T$. (b) CPB's first 3 energy levels computed
  numerically for $E_J / E_C = 0.2$ as a function of the reduced charge $n_G$
  accumulated on the gates. (c) Corresponding quantum capacitance $C_Q$ of the
  ground state of the CPB (relative to the absolute maximum at $n_G = 1 / 2$).
  Vertical dashed lines indicate the working gate charge $n^{\tmop{opt}}_G$ at
  which sensitivity to a small change in $n_G$ is optimum.}
\end{figure}

The Cooper pair box (CPB - see fig. \ref{NRJ_Cq_of_nG}.a)
{\cite{buttiker_zero-current_1987,bouchiat_quantum_1998,nakamura_coherent_1999,vion_manipulating_2002}},
consists of a micron-sized superconducting ``island'' connected to a much
larger superconducting ``reservoir'' through a Josephson junction. The island
can be capacitively coupled to other electrodes to sense the charge that they
carry.

The CPB is characterized by a single degree of freedom: the charge on its
island and its conjugate variable the phase across the junction. Its behavior
results from a competition between charging effects which tend to localize
charge in the island, and the Josephson effect that allows Cooper pairs to
tunnel in and out. Both are quantified by their characteristic energy: (i)
$E_C = (2 e)^2 / 2 C_{\Sigma} $ the charging energy of one Cooper pair, with
$C_{\Sigma} = C_J + \sum_i C_{G, i}$ the sum of the Josephson junction
capacitance $C_J$ and all the capacitances $C_{G, i}$ between the island and
gates $i$; (ii) $E_J = \Phi_0 I_C / 2 \pi$ the Josephson energy coupling two
charge states of the island differing by one Cooper pair, with $\Phi_0 = h / 2 e$ the magnetic flux quantum and $I_C$ the critical current of the junction.

\subsubsection{The CPB Hamiltonian}

The CPB Hamiltonian, written in the basis of charge states $| n \rangle$
corresponding to integer numbers $n$ of Cooper pairs on the island, reads
\begin{equation}
\begin{split}
  H_{\tmop{CPB}} = \underset{n = - \infty}{\overset{+ \infty}{\sum}} \{
  &E_C (\hat{n} - n_G)^2 | n \rangle \langle n| 
  \\
  &- \frac{E_J}{2} (| n + 1
  \rangle \langle n| + | n \rangle \langle n + 1|) \},
  \label{CPB-Hamiltonian}
\end{split}
\end{equation}
where the ``gate charge'' $n_G = \sum n_{G, i} + n^{\sim}_G$ is the sum of all
reduced charges $n_{G, i} = C_{G, i} V_{G, i}  / 2 e$ of the surrounding gate
electrodes at voltages $V_{G, i}$, plus a fluctuating offset charge
$n^{\sim}_G$ of microscopic origin. $H_{\tmop{CPB}}$ can be diagonalized
either numerically by truncating the charge basis, or analytically in the
conjugate phase basis {\cite{cottet_implementation_2002}}. The eigenstates of
the CPB are coherent superpositions of charge states forming 2e-periodic
energy bands $\varepsilon_k (n_G)$, the shape of which depends on the $E_J /
E_C $ ratio. In the so-called charge regime $E_J \ll E_C$ of interest for
electrometry, the CPB states are close to pure charge states and their energy
and other observables depend strongly on $n_G$ (see Fig.
\ref{NRJ_Cq_of_nG}.b).

\subsubsection{Our observable: the quantum capacitance in the charge regime}
\label{Quantum_cap}

As can be seen from the CPB spectrum on Fig. \ref{NRJ_Cq_of_nG}.b, the
degeneracy between two neighbouring charge states is lifted around $n_G = 1 /
2$ by the Josephson coupling. This induces a change of curvature
$\partial_{n_G}^2 \varepsilon_0$ of the CPB ground state. This curvature,
which is plotted in Fig. \ref{NRJ_Cq_of_nG}.c, enters the capacitance to
ground of any particular gate electrode $i$, by a quantity $C^{}_{Q, i} \equiv
- \eta_i C_{G, i} \partial_{n_G}^2 \varepsilon_0$ called the quantum or
Josephson capacitance of the CPB
{\cite{widom_josephson_1984,duty_observation_2005-1,sillanpaa_direct_2005}},
with $\eta_i = C_{G, i}  / C_{\Sigma}$ the gate $i$ lever arm.

The quantum capacitance is obviously higher in the charge regime: For $E_J /
E_C \lesssim 0.1$, a subspace of two charge states accurately describes the
system, and
\begin{equation}
  C^{}_{Q, i} \simeq - \eta_i C_{G, i}  \frac{E_C E_J^2 }{[E_C^2  (1 - 2
  n_G)^2 + E_J^2]^{3 / 2}} \label{Cqi}
\end{equation}
can even exceed $C_{\Sigma}$. It can be used for instance to probe the state
of the CPB charge qubit {\cite{persson_fast_2010}}.

The modulation of $C_Q$ with gate charge $n_G$ also enables charge sensing:
charge fluctuations on a gate can indeed be detected by monitoring the CPB's
effective capacitance, provided that thermal excitations of the CPB at all
$n_G$ are avoided by operating it at $k_B T \ll \min (\varepsilon_1 -
\varepsilon_0) = E_J$. Moreover, because $C^{}_{Q, i}$ can be measured in ac
at relatively high frequency, it is a good observable for high-speed
electrometry.

\subsection{Measurement scheme: Embedding the CPB in a
resonator}\label{CPB+cavity}

\begin{figure}[!th]
  \resizebox{1\columnwidth}{!}{\includegraphics{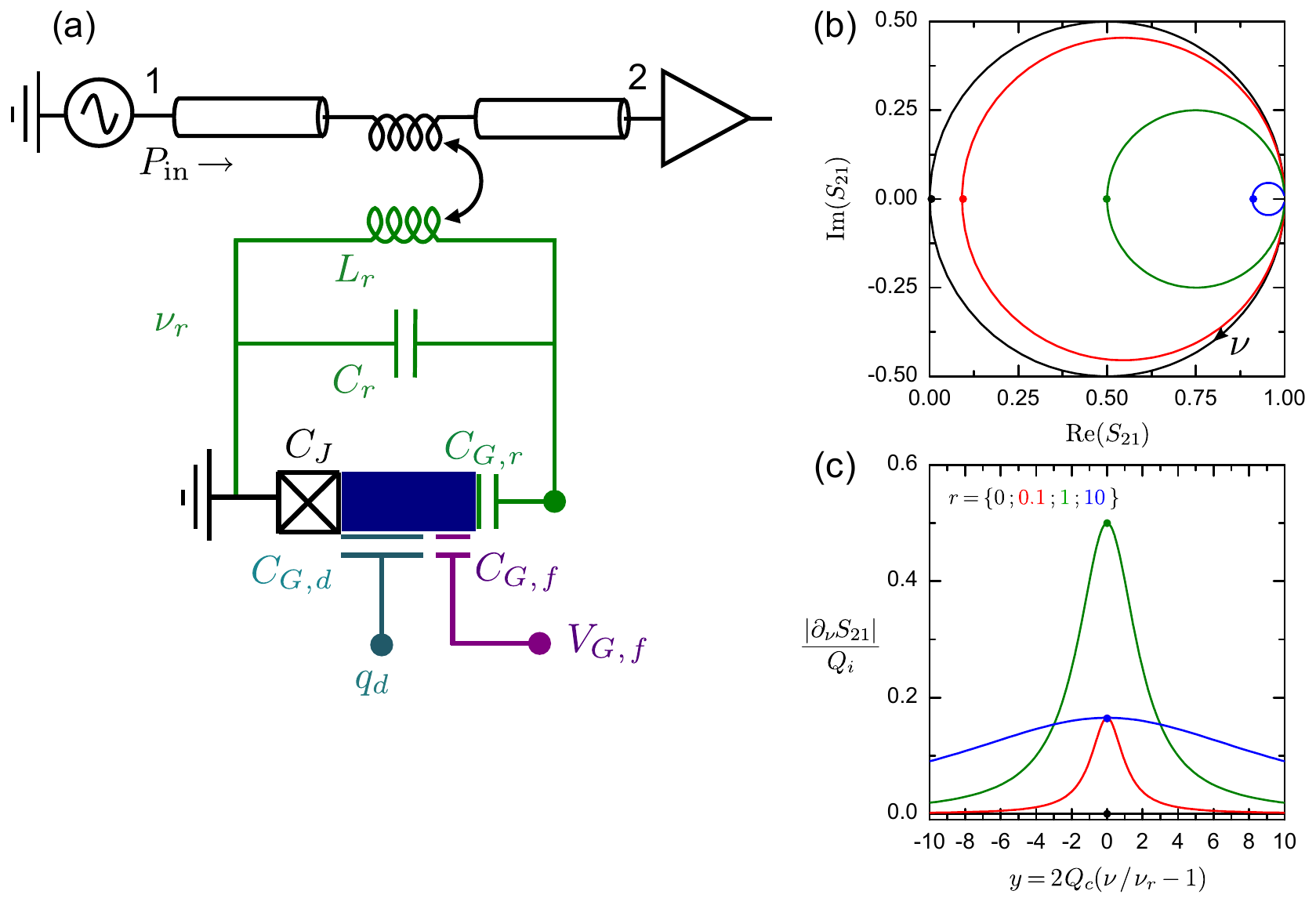}}
  \caption{\label{meas_scheme}Microwave CPB electrometer. (a) The resonator
  embedding the CPB is ``side-coupled'' inductively to a $50 \Omega$
  transmission line along which the $\mu$w excitation propagates and is
  amplified. The CPB quantum capacitance $C_Q$ seen from the $\mu$w gate
  acts as a parallel capacitance that adds up to the bare resonator
  capacitance $C_r$. The charge $q_d$ collected on the detector gate capacitor
  $C_{G, d}$ modifies $C_Q (n_G)$ and shifts the resonator frequency $\nu_r$.
  To maximize the response to charge variations a ``feedback'' gate (labelled
  $f$) is used to maintain $n_G $ at the optimal value for which
  $\partial_{n_G} C_Q$ is maximum. (b) The resonator is probed through its
  $S_{21}$ complex transmission coefficient between ports 1 and 2. When
  varying the frequency $S_{21}$ moves in the complex plane along a circle
  starting from and ending at $\tmop{Re} (S_{21}) = 1$ far away from
  resonance, and passing through a minimum amplitude at resonance (dots); the
  diameter of this circle depends on the ratio $Q_c / Q_i$ of the internal
  losses to the energy leakage towards the line. (c) Linear responsivity $|
  \partial_{\nu} S_{21} |$ of the resonator (see section \ref{rf-response})
  plotted as a function of reduced frequency $y$, for the same set of $r = Q_c
  / Q_i$. Responsivity is maximal at resonance ($y = 0$) and for critical
  coupling ($r = 1$), and increases with $Q_i$.}
\end{figure}

To measure its quantum capacitance variations, the CPB is capacitively coupled
to a superconducting LC resonator of frequency $\nu_r = 1 / 2 \pi \sqrt{L_r
C_r}$ and impedance $Z_r = \sqrt{L_r  / C_r}$ with a coupling capacitance
$C_{G, r}$ (see Fig. \ref{meas_scheme}). According to previous section, the
resonator is thus terminated by an effective tunable capacitance
$C_{\tmop{eff}} = C_{\tmop{geom}, r} + C_{Q, r} (n_G)$ (see Fig.
\ref{NRJ_Cq_of_nG}). Here, $C_{\tmop{geom}, r} = (1 - \eta_{G, r}) C_{G, r}$
is the geometrical capacitance seen from the microwave port of the CPB.
Classically, $C_{\tmop{eff}}$ adds to the resonator capacitance $C_r$ and
displaces its bare frequency $\nu_r$ to $\tilde{\nu}_r (n_G) = \nu_r (1 +
C_{\tmop{eff}} (n_G) / C_r)^{- 1 / 2}$.

The resonator is coupled inductively to the side of an on-chip transmission
line of characteristic impedance $Z_0 = 50 \Omega$ whose complex transmission
$S_{21} (\nu = \omega / 2 \pi) = V^-_2 / V^+_1$ at frequency $\nu$ is measured
close to the resonator frequency $\tilde{\nu}_r$ (see Fig. \ref{meas_scheme}).
The frequency shift of the resonator induced by the CPB modifies the phase and
amplitude of $S_{21}$. The output signal at port 2 is then amplified by a
first amplifier stage whose input noise determines the overall sensitivity of
the setup, as discussed below.

This ``side-coupling'' is inspired from detector arrays for astroparticle and
astronomy, and is used in a variety of applications including arrays of
RF-SETs {\cite{stevenson_multiplexing_2002-1}}, kinetic inductance detectors
(KIDs) {\cite{day_broadband_2003-1}}, SQUIDs {\cite{irwin_microwave_2004}} and
qubits {\cite{schmitt_multiplexed_2014}}. It can provide up to a few
$\tmop{MHz}$ bandwidth and enables multiplexing a large number of detectors on
a single microwave line.

\section{Small signal sensitivity}\label{sensitivity}

We now derive the expressions for the sensitivity $S_e$ of our electrometer.
In sections \ref{dCqdnG} to \ref{SeCq}, we first make a simple calculation
that neglects the nonlinearity of the resonator inherited from its coupling to
the CPB. In section \ref{NL-responsivity} we take this
effect into account, and show how the nonlinearity can be exploited to improve
the sensitivity. We finally define the procedure to find the optimal
parameters.

The electrometer responsivity, defined as the modulus of the $S_{21}$
derivative with respect to $n_G$,
\begin{equation}
\begin{split}
    | \partial_{n_G} S_{21} | = &| \partial_{\nu} S_{21} | \cdot \partial_{C_Q}
  \nu_r \cdot \partial_{n_G} C_{Q, r} \quad 
   \texorpdfstring{\\}{}
   &[V / 2 e / \tmop{input} V]
  \label{rng},
\end{split}
\end{equation}
is simply the product of the variation of $S_{21}$ with frequency, the
variation $\partial_C \nu_r = \partial_{C_Q} \nu_r$ of the resonance frequency
with capacitance, and the variation of $C_Q$ with gate charge. This expression
is valid in the linear regime, when the $\mu$w probe signal represents
small variations $\delta n_G \ll 1$ of $n_G$.

\subsection{Capacitance response \texorpdfstring{$\partial_{n_G} C_{Q,\tmop{rf}}$}{dCqdnG}}
\label{dCqdnG}

The capacitance derivative with respect to charge can be expressed
analytically from Eq. (\ref{Cqi}) for $E_J / E_C \lesssim 0.1$ and small
$\mu$w power:
\begin{equation}
  \partial_{n_G} C_{Q, r} \simeq \eta_{G, r} C_{G, r}  \frac{6 E_C^3 E_J^2 (2
  n_G - 1)}{ [E_C^2 (1 - 2 n_G)^2 + E_J^2]^{5 / 2}} \label{dCqdq} .
\end{equation}
It takes a maximum $\propto \left( E_C  / E_J \right)^2$ at $n^{\tmop{opt}}_G
= 1 / 2 \pm E_J  / 4 E_C$. This value of $n_G$ is the optimal working point,
which is to be adjusted constantly during operation with a DC ``feedback''
gate (see Fig. \ref{meas_scheme}.a).

\subsection{Microwave response \texorpdfstring{$\partial_{C_Q} \nu_r$}{dfdCq} and \texorpdfstring{$|\partial_{\nu}S_{21} |$}{dS21df}}
\label{rf-response}

The second term entering Eq.(\ref{rng}) is written $\partial_{C_Q} \nu_r = -
\nu_r  / 2 C_r $, assuming that the $\mu$w voltage across the resonator
capacitor appears fully on the $\mu$w gate of the CPB (i.e. that the CPB
is effectively placed in parallel with the capacitance of the resonator).

To obtain $| \partial_{\nu} S_{21} |$, we first express the complex
transmission coefficient $S_{21}$ (see Fig. \ref{meas_scheme}.b) as a function
of the reduced frequencies $x = \nu / \nu_r - \nu_r / \nu \approx \nu / \nu_r
- 1$ and $y = 2 Q_c x$. In the side-coupled geometry
\begin{equation}
  S_{21} (x) = 1 - \frac{1}{1 + Q_c / Q_i + 2 i Q_c x} = \frac{1}{1 +
  \frac{1}{r + iy}} \label{S21},
\end{equation}
where $Q_i$ and $Q_c$ are the resonator quality factors due to internal losses
and coupling to the $50 \Omega$ transmission line, respectively, and $r = Q_c 
/ Q_i$.

The internal quality factor can be expressed as $Q^{- 1}_i =
(\kappa_{\tmop{TLS}} + \kappa_{\tmop{rad}} + \kappa_{\tmop{diss}} +
\kappa_{\tmop{mag}} + \kappa_{\tmop{ports}}) / \omega_r$, taking into account
the energy decay rates due to dielectric two level systems (TLS)
{\cite{martinis_decoherence_2005-1}}, radiation in free space (rad), Joule
dissipation (diss), magnetic field induced losses (mag), or losses towards
other electrodes coupled to the resonator (e.g., the gates connected to the
CPB).

The coupling quality factor $Q_c$ is given by the mutual inductance $M_c$
and/or coupling capacitance $C_c$ to the transmission line. For weak coupling
we have $Q_c \approx | Z_{\tmop{coupler}} |^2 / Z_{\tmop{line}} Z_r$ (resp.
$Q_c \approx Z_{\tmop{line}} Z_r  /  | Z_{\tmop{coupler}} |^2$) in the case of
a purely capacitive (resp. inductive) coupling, with $Z_{\tmop{coupler}} = 1 /
j C_c \omega_r$ (resp. $j M_c \omega_r$) the impedance of the coupling element
at the resonator frequency $\omega_r$, and $Z_{\tmop{line}} = Z_0 / 2$ the
line impedance as seen from the coupling point. For a distributed coupling,
this generalizes to $Q_c \approx \omega^2_c / \omega^2_0$ , with $\omega^2_c =
1 / M_c C_c$, with values of $C_c$ and/or $M_c$ that are usually deduced from
numerical simulations.

It is also useful to define the total quality factor $Q^{- 1}_t = Q^{- 1}_c +
Q^{- 1}_i$ that relates the power $P_{\tmop{in}}$ injected at the input port 1
of the transmission line (Fig. \ref{meas_scheme}) to the average number of
photons $\bar{N}$ inside the resonator:
\[ \bar{N} = \frac{P_{\tmop{in}}}{\hbar \omega^2_r} \left[ \frac{Z_0}{Z_r} +
   \frac{2 Q_t}{(1 + r) } \right] \approx \frac{P_{\tmop{in}}}{\hbar
   \omega^2_r} \frac{2 Q_t}{(1 + r) } \quad \tmop{for} Q_t \gg 1. \]
The modulus of the $S_{21}$ variations with respect to a change in the
resonance frequency,
\begin{equation}
  | \partial_{\nu} S_{21} | = \frac{1}{\nu_r}  \frac{2 r Q_i }{(1 + r)^2 +
  y^2} \label{dS21df} \quad [V / \tmop{Hz} / \tmop{input} V],
\end{equation}
has a maximum at the resonance frequency ($y = 0$) and increases with $Q_i$
(see Fig. \ref{meas_scheme}.c). Given some unavoidable internal losses (onto
which one has limited control), the best resonator linear responsivity is
achieved at critical coupling $Q_c = Q_i$ ($r = 1$). However we will see in
section \ref{dS21dfNL} that the result is different when taking the
nonlinearity into account.

\subsection{Linear sensitivity limited by the amplification
chain}\label{Selin}

The detection sensitivity is now determined by the noise sources. The main and
only sample-independent one comes from the $\mu$w amplification chain, the
noise of which is always dominated by the first amplification stage. In this
section where we consider only linear behaviours of all the components, the
first amplification stage is a high electron mobility transistor (HEMT)
amplifier located at 4K. Typical commercially available HEMT amplifiers have
noise temperatures around $T_N = 4 K$ in the frequency range of interest ($[1
- 5] \tmop{GHz}$), corresponding to a noise power $k_B T_N$ per unit
bandwidth. By definition, the sensitivity $S_e$ of the electrometer in $e^- /
\sqrt{\tmop{Hz}}$ is the charge variation inducing a power variation at the
amplifier input equal to $k_B T_N \times 1 \tmop{Hz}$ (signal to noise ratio
of 1 in a bandwidth of 1Hz). It is thus the voltage noise spectral density
$\sqrt{k_B T_N Z_0}$ divided by the electrometer responsivity, that is
\begin{equation}
  S_e = \sqrt{k_B T_N / (P_{\tmop{in}} | \partial_{n_G} S_{21} |^2 / 4)}
  \label{Se-generic}
\end{equation}
with $P_{\tmop{in}}$ the power at the sample input. Combining Eq.
(\ref{dCqdq}) and (\ref{dS21df}) into Eq. (\ref{rng}) yields the setup limited
charge sensitivity in the linear regime at low $\mu$w power and optimal
$n_G$:
\begin{equation}
  \begin{array}{lll}
    S^{\tmop{lin}}_e & = & c_0 \sqrt{\frac{k_B T_N}{P_{\tmop{in}}}}  \frac{(1
    + r)}{Q_t}  \frac{C_r}{\eta_{G, r} C_{G, r}}  \left( \frac{E_J}{E_C}
    \right)^2\\
    & = & \frac{c_0}{4} \sqrt{\frac{k_B T_N}{P_{\tmop{in}}}}  \frac{(1 +
    r)}{Q_t}  \left( \frac{E_J}{\hbar g} \right)^2  \frac{\hbar \omega_r}{E_C}
    \quad \left[ e^- / \sqrt{\tmop{Hz}} \right],
  \end{array}  \label{Sqlin}
\end{equation}
where $c_0 = 25 \sqrt{5} / 48 \simeq 1$ and $\hbar g = \sqrt{\frac{\eta_{G, r}
C_{G, r} }{4 C_r}} \sqrt{\hbar \omega_r E_C}$ expresses the capacitive
coupling between the CPB and the resonator in terms of energy (or frequency),
as detailed further in section \ref{hamiltonian-analysis}.

Optimizing our electrometer consists in minimizing $S^{\tmop{lin}}_e$ by
adjusting all {\tmem{independent}} parameters. At first glance, parameters in
(\ref{Sqlin}) could seem independent, though it is not the case.
$P_{\tmop{in}}$ is constrained below a maximum value due to two
nonlinearities: the nonlinearity of the quantum capacitance response
($\partial_{n_G} C_{Q, \mu}$) presented in the next section, and the
nonlinearity of the resonator response ($| \partial_{\nu} S_{21} |$) inherited
from its coupling to the CPB, treated in section \ref{hamiltonian-analysis}.

\subsection{Smoothing of the quantum capacitance by the microwave
signal}
\label{SeCq}

Since the CPB's quantum capacitance is nonlinear in $n_G$, the $\mu$w gate
charge $\delta n_G = \eta_{G, r}  \sqrt{\pi Z_r / R_Q} \sqrt{\bar{N}} \hbar
\omega_r  / E_C$, with $R_Q = h / 4 e^2$ the superconducting quantum of
impedance, has to be small compared to the period ($\delta n_G \ll 1$).
Otherwise, the sharp $C_Q$ variations around $n^{\tmop{opt}}_G$ responsible
for the high sensitivity of the electrometer (see Fig. \ref{NRJ_Cq_of_nG}.c)
is smoothed and the sensitivity degraded. This smoothing is calculated here in
a semi-classical way. We define the average capacitance
\[ \overline{C_{Q, r}} = \frac{1}{2 \pi} \int^{2 \pi}_0 C_{Q, r} \left( n_{G,
   \tmop{dc}} + \frac{\delta n_G}{2} \cos \theta \right) d \theta \]
probed by the $\mu$w tone as the integral of the quantum capacitance over
a period of the microwave, assuming a linear resonator response.

\begin{figure}[!t]
  \resizebox{1\columnwidth}{!}{\includegraphics{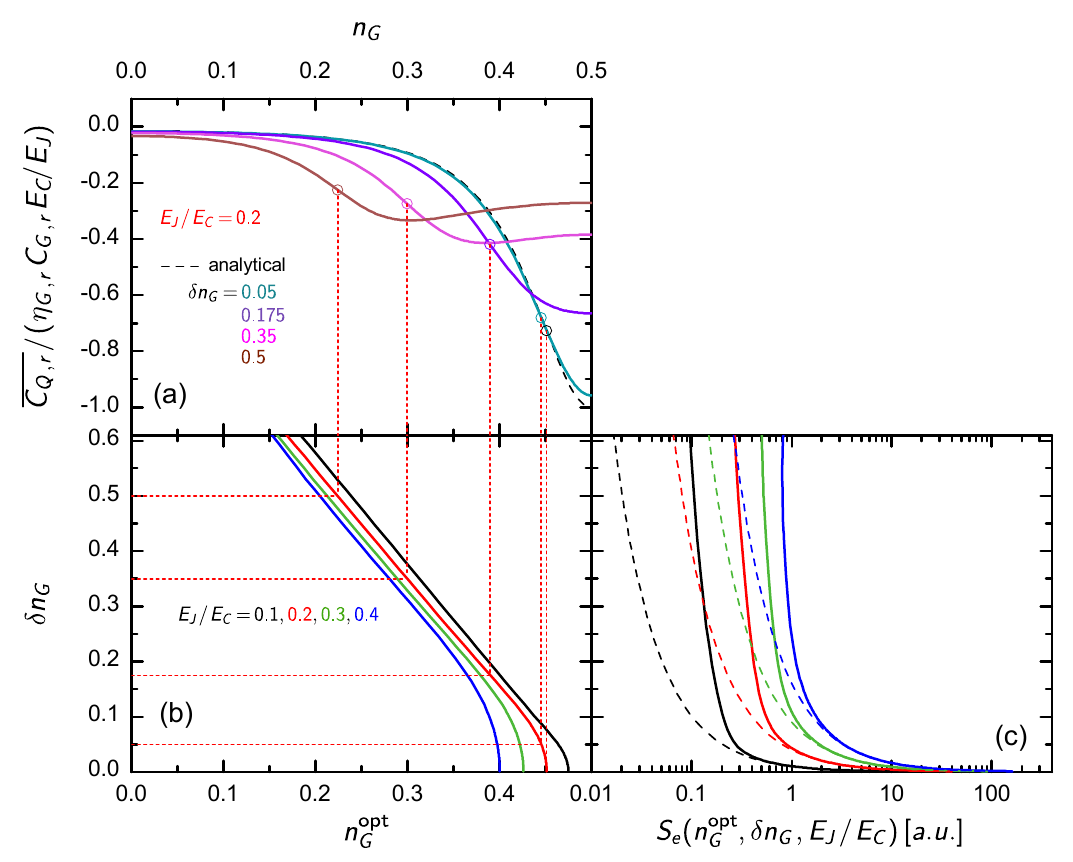}}
  \caption{\label{sens_vs_pow}Smoothing of $C_Q$ with $\mu$w probe
  amplitude ($\propto \delta n_G$) and degradation of sensitivity $S_e$. (a)
  Smoothed quantum capacitance $\overline{C_{Q, r}}$ as a function of $n_G$
  for $E_J / E_C = 0.2$ and several $\mu$w excitations $\delta n_G$. Dots
  indicate the resulting optimal bias $n^{\tmop{opt}}_G (\delta n_G)$, at
  which charge responsivity $\partial_{n_G} C_{Q, r}$ is maximum. (b)
  Variation of $n^{\tmop{opt}}_G$ with $\delta n_G$ (swapped axes) at four
  values of $E_J / E_C$. (c) Sensitivity $S_e$ versus $\mu$w amplitude
  $\delta n_G$ at $n^{\tmop{opt}}_G (\delta n_G)$ (continuous lines), degraded
  compared to the linear sensitivity (dashed lines) of Eq. (\ref{Sqlin}).}
\end{figure}

Figure \ref{sens_vs_pow}.a shows the result of this averaging for several
values of the $\mu$w excitation $\delta n_G \lesssim 1 / 2$ . Note that
the optimal working point $n_G^{\tmop{opt}}$ now depends on power
$P_{\tmop{in}}$ and has to be computed numerically (Fig. \ref{sens_vs_pow}.b).
The resulting $\mu$w averaged sensitivity $S_e (n^{\tmop{opt}}_G)$
degrades at high $\delta n_G$ as shown in Fig. \ref{sens_vs_pow}.c. This
degradation can be kept below a factor two by limiting $\delta n_G$ below
$0.1$.

\subsection{Nonlinearity of the resonator response}\label{NL-responsivity}

\subsubsection{The CPB-resonator Hamiltonian}\label{hamiltonian-analysis}

Up to now we have forgotten the nonlinearity of the resonator inherited from
the very nonlinear CPB device coupled to it. We will show that this effect can
be already important for only a few photons in the resonator and that it
modifies strongly the sensitivity dependence (\ref{Sqlin}) on input power,
quality factors and resonator-CPB coupling. This nonlinearity is calculated
below with a full quantum treatment of the CPB-resonator system.

In the circuit quantum electrodynamics language, the total Hamiltonian of the
system reads {\cite{blais_cavity_2004,koch_charge-insensitive_2007}}
\begin{eqnarray*}
  H_{\tmop{tot}} & = & H_r + H_{\tmop{CPB}} + H_c
\end{eqnarray*}
with
\begin{equation}
  \begin{array}{lll}
    H_r & = & \hbar \omega_r  \left( a^{\dag} a + \frac{1}{2} \right),\\
    H_{\tmop{CPB}} & = & \sum_k \varepsilon_k  | k \rangle \langle k|,\\
    H_c & = & 2 \hbar g \hat{n} (a + a^{\dag}),
  \end{array} \label{Htot}
\end{equation}
where $a$ and $a^{\dag}$ are the photon annihilation and creation operators of
the bare resonator, and $\hbar g = e V_{\tmop{rms}} C_{G, r} / C_{\Sigma} $ is
the CPB-resonator coupling factor with $V_{\tmop{rms}} = \sqrt{\hbar \omega_r
/ 2 C_r}$ the rms voltage on the $\mu$w gate of the CPB produced by zero
point fluctuations. Here, the field amplitude $(a + a^{\dag})$ in the coupling
term reflects the $\mu$w gate charge amplitude $\delta n_G =
\sqrt{\bar{N}} 2 \hbar g / E_C$ entering the CPB's charge term $E_C (\hat{n} -
n_G)^2$ in Eq. (\ref{CPB-Hamiltonian}).

Expressing the operator $\begin{array}{l}
  \hat{n} = \sum_{k, l} \langle k| \hat{n} | l \rangle  | k \rangle \langle l|
\end{array}$ in the CPB eigenbasis using $\begin{array}{lll}
  \langle k| \hat{n} | l \rangle & = & - i \int^{^{\pi}}_{- \pi} d \theta
  \varphi^{\ast}_k (\theta) \partial_{\theta} \varphi_l (\theta) 
\end{array}$and $\begin{array}{lll}
  \varphi_k (\theta) & = & \langle \theta | \nobracket k \rangle
\end{array}$, where $\varphi_k (\theta)$ is the CPB wavefunction in the
conjugate phase representation {\cite{cottet_implementation_2002}}, allows us
to diagonalize $H_{\tmop{tot}}$ exactly in the tensorial product of the two
eigenbases.

Since our CPB is operated in the charge regime $E_J / E_C \ll 1$ at
$n^{\tmop{opt}}_G$ close to $1 / 2$, one could think that it is modeled
sufficiently accurately by the two level system (TLS) made of the $\{ | n = 0
\rangle, | n = 1 \rangle \}$ charge states at $n_G = 1 / 2$. In this
approximation $H_{\tmop{tot}}$ takes the Jaynes-Cumings form (eq. 16 in
{\cite{blais_cavity_2004}} with $\theta = \pi / 2$ )\tmcolor{orange}{}
\[ \begin{array}{lll}
     H_{\tmop{JC}} & = & \hbar \omega_r  \left( a^{\dag} a + \frac{1}{2}
     \right) + \frac{\hbar \Omega}{2} \sigma_Z + \hbar g (\sigma^+ + \sigma^-)
     (a + a^{\dag}),
   \end{array} \]
with $\hbar \Omega \equiv \varepsilon_1 - \varepsilon_0$ the TLS transition
energy, and $\sigma^+$, $\sigma^-$ and $\sigma_Z$ the TLS raising, lowering,
and z Pauli operators, respectively. Neglecting the fast oscillating terms
$\sigma^+ a^{\dag}$ and $\sigma^- a$ that do not conserve the number of
excitations and taking into account that the CPB transition and the resonator
frequencies are very different ($\Delta \equiv \Omega - \omega_r \gg g$),
$H_{\tmop{JC}}$ can be solved analytically in the so-called rotating wave
approximation (RWA), which yield a dispersive hamiltonian
{\cite{koch_charge-insensitive_2007}}
\[ H_{\tmop{disp}} = \hbar (\omega_r + \chi \sigma_z) a^{\dag} a + \hbar
   (\Omega - \chi) \sigma^+ \sigma^- + \frac{\hbar K_{\tmop{TLS}}}{2} a^{\dag
   2} a^2 \]
exhibiting a CPB state-dependent resonator shift $\mp \chi = \mp g^2 / \Delta$
and a Kerr nonlinearity per photon {\cite{nigg_black-box_2012}}
\begin{equation}
  K_{\tmop{TLS}} = 2 \Delta (g / \Delta)^4 . \label{KTLS}
\end{equation}
However, within the relevant parameter window for electrometry, solving
numerically $H_{\tmop{tot}}$ (or $H_{\tmop{JC}}$ keeping all terms) yields
very different results from those obtained for $H_{\tmop{disp}}$ in the RWA,
the nonlinearity at low photon numbers being for instance three times larger.

Consequently, we solve Eq. (\ref{Htot}) numerically by truncating the CPB and
resonator Hilbert spaces to three eigenstates and $N_{\tmop{tot}} = 200$ Fock
states, respectively {\cite{main_note_1}}. From the eigenspectrum, we select
the energies $E_{0, N}$ of the eigenstates $| 0, N \rangle$ corresponding to
the CPB being mostly in its ground state $|0 \rangle$ hybridized with
the resonator mostly in Fock state $| N \rangle$. We then plot in Fig. 4 the
transitions frequencies $\Delta \omega_{N, \tmop{rel}} \equiv (\omega_N -
\omega_r) / \omega_r = (E_{0, N + 1} - E_{0, N}) / \hbar \omega_r - 1$ between
two successive $| N \rangle \tmop{states}$, relative to the bare resonator
frequency $\omega_r$, together with the results obtained within the RWA
(dashed lines).

\begin{figure}[!t]
  \resizebox{1\columnwidth}{!}{\includegraphics{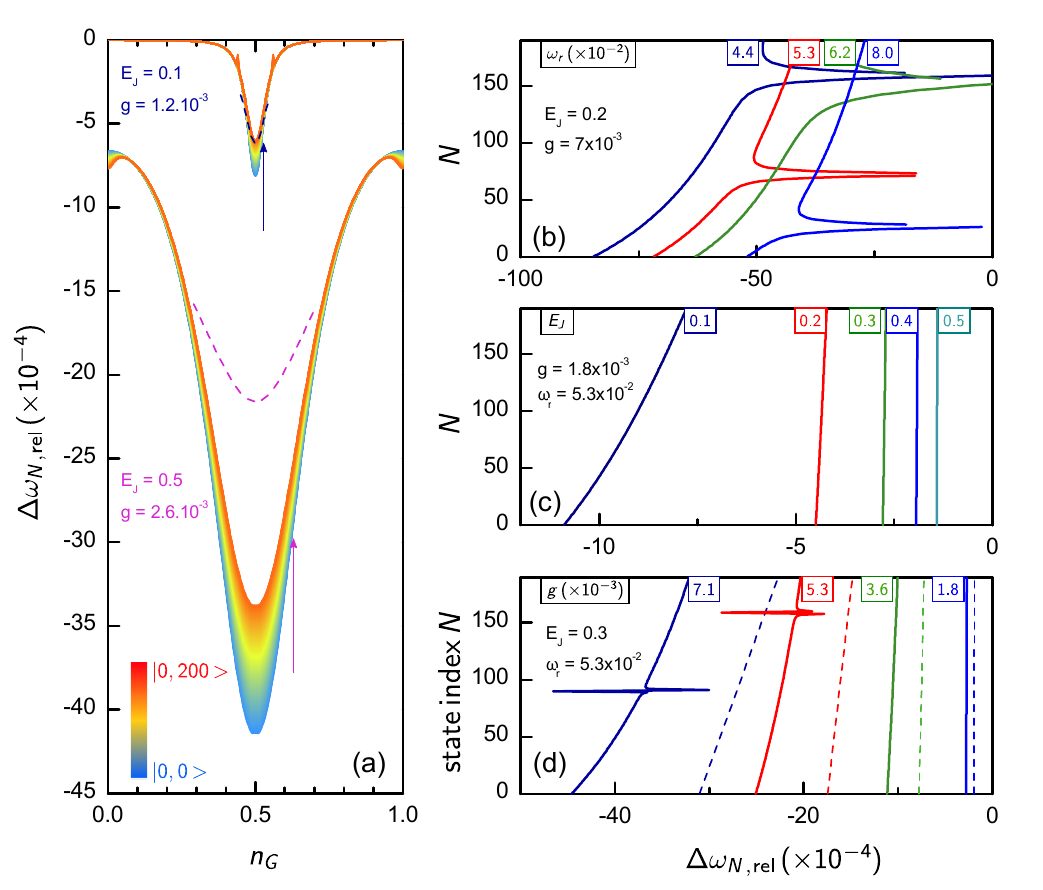}}
  \caption{\label{CPB-cav-spectrum}Resonator's relative transition frequencies
  $\Delta \omega_{N, \tmop{rel}}$ (the CPB being in its ground state)
  obtained by numerical diagonalization of the CPB-resonator Hamiltonian
  (solid lines), or using the Hamiltonian $H_{\tmop{disp}}$ (RWA - dashed
  lines). All parameters $E_J$, $\omega_r$, and $g$ are expressed in units of
  $E_C$. (a) Plot as a function of $n_G$ from $N = 0$ to $N = 200$ (color
  scale) for two pairs of \{$E_J$, $g$\} values. The $n^{\tmop{opt}}_G$
  positions are indicated by arrows. The RWA is plotted at $N = 0$ only.
  (b,c,d) Plots as a function of the Fock state index $N$, at
  $n^{\tmop{opt}}_G$, and for different bare resonator frequencies (b),
  Josephson energies (c), and CPB-resonator couplings $g$ (d), keeping all
  other parameters fixed.}
\end{figure}

The dependence on $n_G$ is shown on Fig. 4.a for two pairs of \{$E_J$, $g$\}
values. The resonator nonlinearity shows up as a dispersion of $\Delta
\omega_{N, \tmop{rel}}$ with Fock index $N$, with a maximum spread at $n_G = 1
/ 2$. This shift of $\Delta \omega_{N, \tmop{rel}}$ with $N$ is plotted in
Fig. \ref{CPB-cav-spectrum}.b-d (swapped axis) at
$n^{\tmop{opt}}_G$ for different sets of parameters $\{ \omega_r, E_J, g \}$.
The spectrum and its dependence on $N$ differ significantly from the RWA
prediction, as seen by comparing the dashed lines with the solid lines in
insets a and d of Fig. \ref{CPB-cav-spectrum}. Besides, we also observe peaks at certain $N$
values, corresponding to resonant transitions $|0, N > \leftrightharpoons |1,
N - m >$involving a coherent exchange of $m$ photons. Their existence depends
on the coupling strength $g$ as shown in fig.
\ref{CPB-cav-spectrum}.d, and the specific $N$ at which they
occur depends on $\omega_r$ and $E_J$ (on the commensurability of the $|0, N
>$and $|1, N >$ ladders), as shown in Fig.
\ref{CPB-cav-spectrum}.b. Such resonance have of course to be
avoided by design for our electrometry application.

These results illustrate how the RWA fails to describe our system, in
particular when $g$ becomes non negligible with respect to $\omega_r$, and
show that the full complexity of the Hamiltonian has to be taken into account
to design properly a $\mu$w-CPB electrometer.

\subsubsection{Nonlinear corrections and \texorpdfstring{\\}{} boost of the charge sensitivity \texorpdfstring{$S_e$}{Se}}
\label{dS21dfNL}

To incorporate easily the nonlinearity computed above in our calculation of
the charge sensitivity, and although this nonlinearity slightly departs from a
pure Kerr behavior, we choose to approximate it by a Kerr nonlinearity with
the Kerr constant $K_0$ that linearizes our numerical results at small $N$:
$\omega_N = \omega_0 + N K_0$.

We now calculate numerically the nonlinear corrections to $| \partial_{\nu}
S_{21} |$ entering the responsivity $| \partial_{n_G} S_{21} |$ in Eq. (\ref{rng}), and how they affect the sensitivity $S_e$ in Eq.
(\ref{Sqlin}). We model the nonlinear resonator by the Hamiltonian
\[ H_{\tmop{NL}} = \hbar \omega_0 a^{\dag} a + \hbar \frac{K_0}{2} a^{\dag 2}
   a^2 \]
with the new frequency $\omega_0$ and Kerr coefficient $K_0$ taken at
$n^{\tmop{opt}}_G$. Following {\cite{yurke_performance_2006}} with the
convention $\beta_{\tmop{re}} = \beta e^{i \omega t} + c.c.$ for all the real
fields $\beta_{\tmop{re}}$, the stationary equation of motion within the
monochromatic approximation at the driving frequency $\omega$ reads
\[ \alpha \left[ \frac{\kappa_t}{2} + i (\omega - \omega_0 - K_0  | \alpha |
   ^2) \right] = i \alpha_{\tmop{in}}  \sqrt{\frac{\kappa_c}{2}}, \]
where $\kappa_c = \omega_0 / Q_C$, $\kappa_i = \omega_0 / Q_i$ and $\kappa_t =
\kappa_c + \kappa_i$ are the modified coupling, internal, and total energy
decay rates, and $\alpha_{\tmop{in}}$ (resp. $\alpha$) is the reduced complex
amplitude of the incident field from port 1 towards the resonator (resp.
internal field inside the resonator), expressed in square root of photons per
seconds (resp. square root of photons).

\begin{figure}[!bt]
  \resizebox{1\columnwidth}{!}{\includegraphics{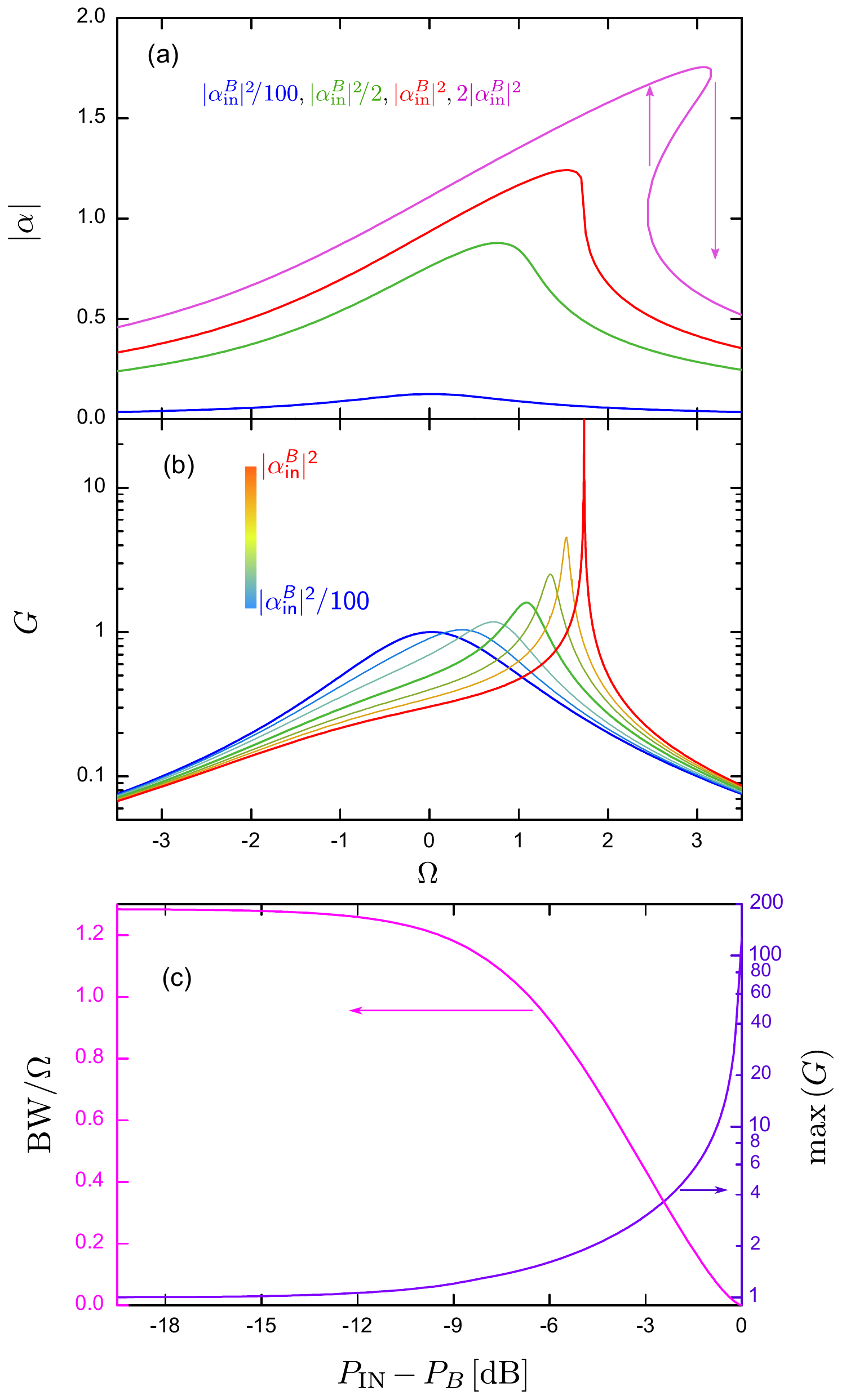}}
  \caption{\label{NLfield+dS21df}Non linear dynamics of the Kerr resonator in
  the side-coupled geometry as a function of the input power. (a)
  Intra-resonator field amplitude $| \alpha |$ as a function of reduced
  frequency $\Omega$ for several input photons rates $| \alpha_{\tmop{in}}^{}
  |^2$ below and above bifurcation. (b) Responsivity $| \partial_{\nu} S_{21}
  |$ versus frequency $\Omega$ at several $| \alpha_{\tmop{in}}^{} |^2$, from
  the linear regime (blue curve) to the bifurcation (red curve), normalized to
  the linear responsivity at $\Omega = 0$. (c) Kerr resonator gain and gain
  bandwidth in $\Omega$ units versus input power relative to bifurcation
  power.}
\end{figure}

This third order polynomial equation has 1 or 3 real solutions depending on
$| \alpha_{\tmop{in}} |$ and on the reduced frequency $\Omega \equiv 2 (\omega
- \omega_0) / \kappa_t$. The bifurcation from 1 to 3 solutions occurs at
$\left\{ \Omega^B = \sqrt{3}, | \alpha_{\tmop{in}}^B |^2 = \kappa_t^3 / \left(
3 \sqrt{3} \kappa_c  | K_0 | \right) \right\}$. The internal amplitudes are
plotted on Fig. \ref{NLfield+dS21df}.a as a function of $\Omega$ for various
input powers $P_{\tmop{in}} = | \alpha_{\tmop{in}}^{} |^2 \hbar \omega_0$.
Above bifurcation, one of the solutions is metastable, and the internal field
is hysteretic (arrows on the figure) and cannot be exploited for continuous
detection. Our electrometer will thus be operated with the resonator below
bifurcation.

Then, the output microwave field
\[ \alpha_{\tmop{out}} \simeq \alpha_{\tmop{in}} +_{} i
   \sqrt{\frac{\kappa_c}{2}} \alpha \]
on port 2 results from the interference between the incident field and the
field re-emitted by the resonator to the right direction, which yields now an
$\alpha -$dependent transmission coefficient
\[ S_{21} = 1 - \frac{\kappa_c}{\kappa_t + 2 i (\omega - \omega_0 - K_0 |
   \alpha |^2)} . \]
The corresponding responsivity normalized to Eq. (\ref{dS21df}) for a
comparison with the linear case, $G = | \partial_{\nu} S_{21} | / |
\partial_{\nu} S_{21} (\alpha_{\tmop{in}} = 0) |$, is plotted in Fig.
\ref{NLfield+dS21df}.b as a function of frequency for several input powers up
to the bifurcation. As bifurcation is approached, an enhancement of the
responsivity is clearly observed ($G > 1$), while the responsivity bandwidth
is reduced {\cite{vijay_invited_2009}}. The gain in responsivity as well as
its corresponding -3dB gain bandwidth $\tmop{BW}$ are plotted on Fig.
\ref{NLfield+dS21df}.c as a function of the incident power.

\begin{figure*}[!t]
  \center
  \resizebox{2\columnwidth}{!}{\includegraphics[width=\textwidth]{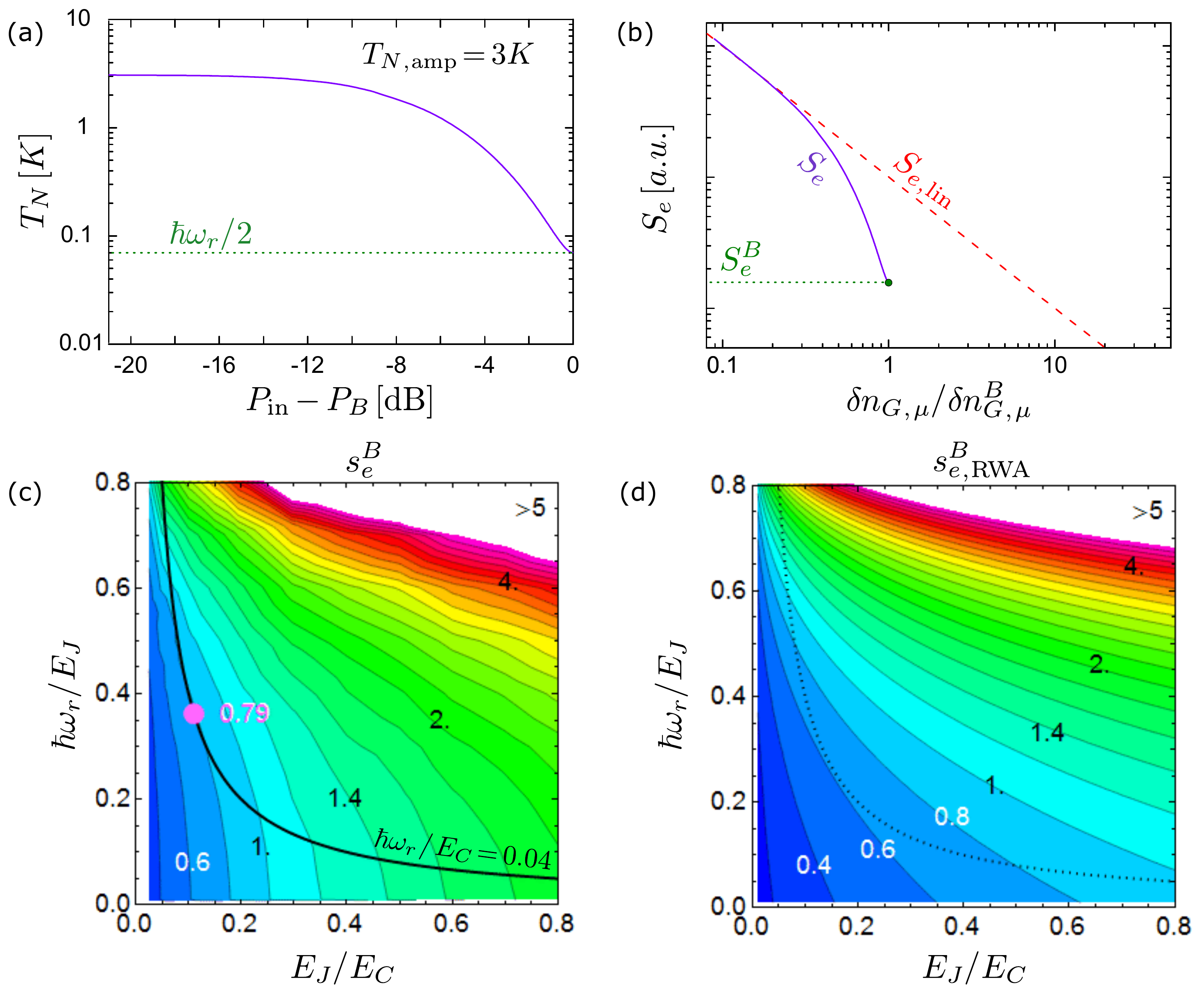}}
  \caption{\label{NLsensitivity}Consequences of parametric amplification on
  the electrometer sensitivity, and optimal choice of parameters assuming a
  noise temperature $T_N = 3 K$ for the HEMT. (a) Effective noise temperature
  versus input power up to bifurcation. The lower bound of $T_N$ reached at
  bifurcation is given by the quantum limit, i.e. $\hbar \omega_r / 2 k_B
  \approx 70 \tmop{mK}$ for a 4 GHz resonator. (b) Reduced sensitivity $s_{e,
  \tmop{RWA}}$ (violet line) and corresponding linear approximation $s^{}_{e,
  \tmop{lin}}$ derived in section \ref{Selin} as a function of the reduced
  $\mu$w gate charge relative to bifurcation. c) Exact reduced
  sensitivities $s_e^B$ as a function of $E_J / E_C$ and $\hbar \omega_r /
  E_J$. The solid line $\hbar \omega_r / E_C = 0.04$ corresponds to the
  constraints $\hbar \omega_r / k_B = 10 T = 120 \tmop{mK}$ and $E_C / k_B = 3
  K$. The magenta point indicates the best (minimum) sensitivity and fixes the
  remaining parameter $E_J / k_B = 0.3 K$. d) Corresponding RWA approximation
  of the sensitivity $s_{e, \tmop{RWA}}^B$.}
\end{figure*}

An interesting property of this non linear resonator is that it lowers the
effective noise temperature of the amplification chain. Indeed, with a
frequency above 4GHz, the superconducting resonator operated in a dilution
refrigerator at a temperature of about 20mK is cold enough to be in the
quantum regime with very little photon noise. More quantitatively, its noise
temperature $T_{N, r} = \hbar \omega_r / 2 k_B$ is much lower than that of the
following HEMT amplifier, and the effective noise temperature of the
amplification chain becomes $T_N = T_{N, r} + T_{N, \tmop{HEMT}} / G^2$. As
the microwave power $P_{\tmop{in}}$ and consequently $\delta n_{G, \mu} = 2
\hbar g \alpha / E_C$ are increased up to bifurcation, $T_N$ decreases as
shown in fig. \ref{NLsensitivity}.a which correlatively improves the
sensitivity. By noting that the dependence of $G$ on $P_{\tmop{in}} - P_B$ is
universal we can infer that the sensitivity has a universal dependence on
$\delta n_G$ (or $\bar{N}$) upon approaching the bifurcation, which we show on
Fig. \ref{NLsensitivity}.b.

Consequently, we choose to operate our electrometer at an input power
$P_{\tmop{in}} \sim P_B$ close to bifurcation. We also choose the
$n^{\tmop{opt}}_G$ working point at which the slope $\partial_q C_Q$ is
negative such that a detected charge $q$ will displace $n_G$ further away
from 1/2, lowering $K_0$ and thus moving the system further away from
bifurcation. Assuming a large quality factor $Q_t$, we thus obtain a quantum
limited charge sensitivity at bifurcation
\begin{align}
  S^B_e (\omega_r, E_J, E_C, g, P_B) = \frac{c_0}{4 \sqrt{2}} \frac{(1 +
  r)}{Q_t}  \frac{\hbar \omega_r}{E_C} \left( \frac{E_J}{\hbar g} \right)^2
  \sqrt{\frac{\hbar \omega_r}{P_B}}
\end{align}
with $P_B = \hbar (1 + r) \omega_r^3 / \left( 3 \sqrt{3} Q^2_t  | K_0 |
\right)$.

The remaining parameter to expand is $K_0$. To do so we use the RWA as an
analytical guideline, even though we have shown in section
\ref{hamiltonian-analysis} that this simplification does not give quantitative
results. Using $K_{\tmop{TLS}}$ from Eq. (\ref{KTLS}) for $K_0$ yields $P_B
\propto g^{- 4}$ and a much simpler form for the sensitivity,
\begin{equation}
\begin{split}
  S^B_{e, \tmop{RWA}} ( & \frac{\hbar \omega_r}{E_J}, \frac{E_J}{E_C}, E_C )
  \\ 
  \thickapprox  &\sqrt{\frac{1 + r}{2}}  \left( 1 - \frac{\hbar
  \omega_r}{E_J} \right)^{- 3 / 2} \sqrt{\frac{E_J}{E_C}} \sqrt{
  \frac{\hbar}{E_C}}, \label{SeRWA}    
\end{split}
\end{equation}
which strikingly no longer depends on $g$ and $P_B$. This can be understood in
the following way: as $g$ increases, better responsivity is achieved while at
the same time the highest allowed working power decreases due to nonlinearity,
hence lowering the signal to noise ratio; those two effects compensate exactly
within the RWA.

We then check numerically that the independence of $S^B_e$ on $g$ and $P_B$
remains true within a few percents for the exact Hamiltonian (\ref{Htot}) and
for all parameter sets satisfying $g / \omega_r \in [0.008 - 0.2]$, provided
$m$-photons resonant exchange processes are avoided. This means that the
sensitivity at bifurcation takes the reduced form $s^B_e = S^B_e / \left(
\sqrt{(1 + r) \hbar / 2 E_C} \right)$ inspired from Eq. (\ref{SeRWA}),

\noindent which
depends only on $\hbar \omega_r / E_J$ and $E_J / E_C$, and that $E_C$ should
be maximized.

We plot $s_e^B (\hbar \omega_r / E_J, E_J / E_C)$ in fig.
\ref{NLsensitivity}.c as well as $s_{e, \tmop{RWA}}^B$ in fig.
\ref{NLsensitivity}.d for comparison. The reduced sensitivity $s^B_e$ is found
to improve at low $\hbar \omega_r / E_J $ and low $E_J / E_C$. However the
optimization is constrained: $E_C$ can be increased only up to a maximum value
that depends on the technology used for the CPB fabrication, and $\omega_r$
cannot be lower than 5-$_{} 10 k_B T / \hbar$ (with $T$ the base temperature
of the electrometer) to keep the resonator in the quantum regime, which
altogether defines a constant product for the remaining variables $(\hbar
\omega_r / E_J) \times (E_J / E_C)$. The optimal sensitivity is thus found at
the minimum of $s^B_e$ along a hyperbola, as the one shown in fig.
\ref{NLsensitivity}.b. for $\hbar \omega_r / k_B = 0.12 K$ and $E_C / k_B = 3
K$, i.e. values that can be reached in many laboratories. The corresponding
optimum sensitivity is found on the graph at $E_J / k_B = 0.3 K$.

To summarize, the sensitivity of the electrometer that we propose will be
optimum when choosing a resonator having any characteristic impedance, the
lowest frequency $\omega_r$ above $5 - 10 k_B T / \hbar$ and $Q$ factors $1000
< Q_c \ll Q_i$ ($r \ll 1$) contrary to what was found from Eq. (\ref{Sqlin}),
when maximizing the charging energy $E_C$, and operating at an input power
$P_{\tmop{in}} \lesssim P_B$ slightly below the resonator bifurcation whatever
the CPB-resonator coupling $g$ is; the only remaining free parameter $E_J$
will finally be optimized according to the procedure described in fig.
\ref{NLsensitivity}.c.

\section{Offset charge noise and quasiparticle poisoning}

We now discuss two problems often encountered when operating a CPB: offset
gate charge noise of microscopic origin, and tunneling of residual
quasiparticles in and out of the CPB island. We present briefly these issues
and describe how to deal with them in practice in an actual electrometer.

\subsection{Offset gate charge noise}

Offset charge refers to randomly fluctuating charged defects in the vicinity
of the CPB island, which induce net gate charge variations (noise). Charge
noise in CPBs has been found to come mainly from oxide defects either inside
{\cite{simmonds_decoherence_2004,lisenfeld_observation_2015}} or close to
{\cite{astafiev_quantum_2004,ithier_decoherence_2005}} the Josephson junction,
that is from locations where the defects are more strongly coupled to the CPB
electric dipole. The magnitude of charge noise can be reduced by improving the
quality of the junction dielectric {\cite{oh_elimination_2006}}, though no
definitive cure has been found yet and the subject is still an active field of
research {\cite{dunsworth_characterization_2017}}.

The spectral density of this noise is known to have a $1 / f^{\alpha}$
dependence {\cite{schoelkopf_radio-frequency_1998}} with $\alpha \backsimeq
1$, due to a collection of widely distributed two level systems
{\cite{shnirman_low-_2005-1,muller_towards_2017}}, which result in a
continuous slow drift of $n^{}_G$. This gives rise to two issues for the
electrometer: (i) its sensitivity is degraded at low frequency due to charge
noise, up to a frequency called the $1 / f$ corner, above which noise is
dominated by $k_B T_N$ as discussed in section \ref{dS21dfNL}. (ii) The
sensitivity reaches its floor $S^B_e$ above the $1 / f$ corner, only on the
condition that $n_G$ does not drift significantly away from
$n_G^{\tmop{opt}}$.

An efficient and easy way to compensate for the slow $n^{}_G$ drift away from
$n_G^{\tmop{opt}}$ is to adapt constantly the voltage applied to a feedback
gate (see Fig. \ref{meas_scheme} and \ref{even-odd}), in order to maintain a
constant quantum capacitance, i.e. maintain the amplitude and phase of the
microwave signal transmitted through the measuring line. This has both the
advantages of keeping the optimal responsivity, and linearizing the
electrometer response to a charge change.

The sensitivity degradation below the $1 / f$ corner can also be partially
cured by coupling a number $N$ of identical CPB-resonator electrometers to the
same device, which yields a $\sqrt{N}$ gain on sensitivity, assuming
uncorrelated charge noises for the different CPBs. Note that the $\mu$w
design proposed here conveniently enables frequency multiplexing of all the
resonators by a single microwave line.

\subsection{Quasiparticle poisoning}

\begin{figure}[!t]
  \resizebox{1\columnwidth}{!}{\includegraphics{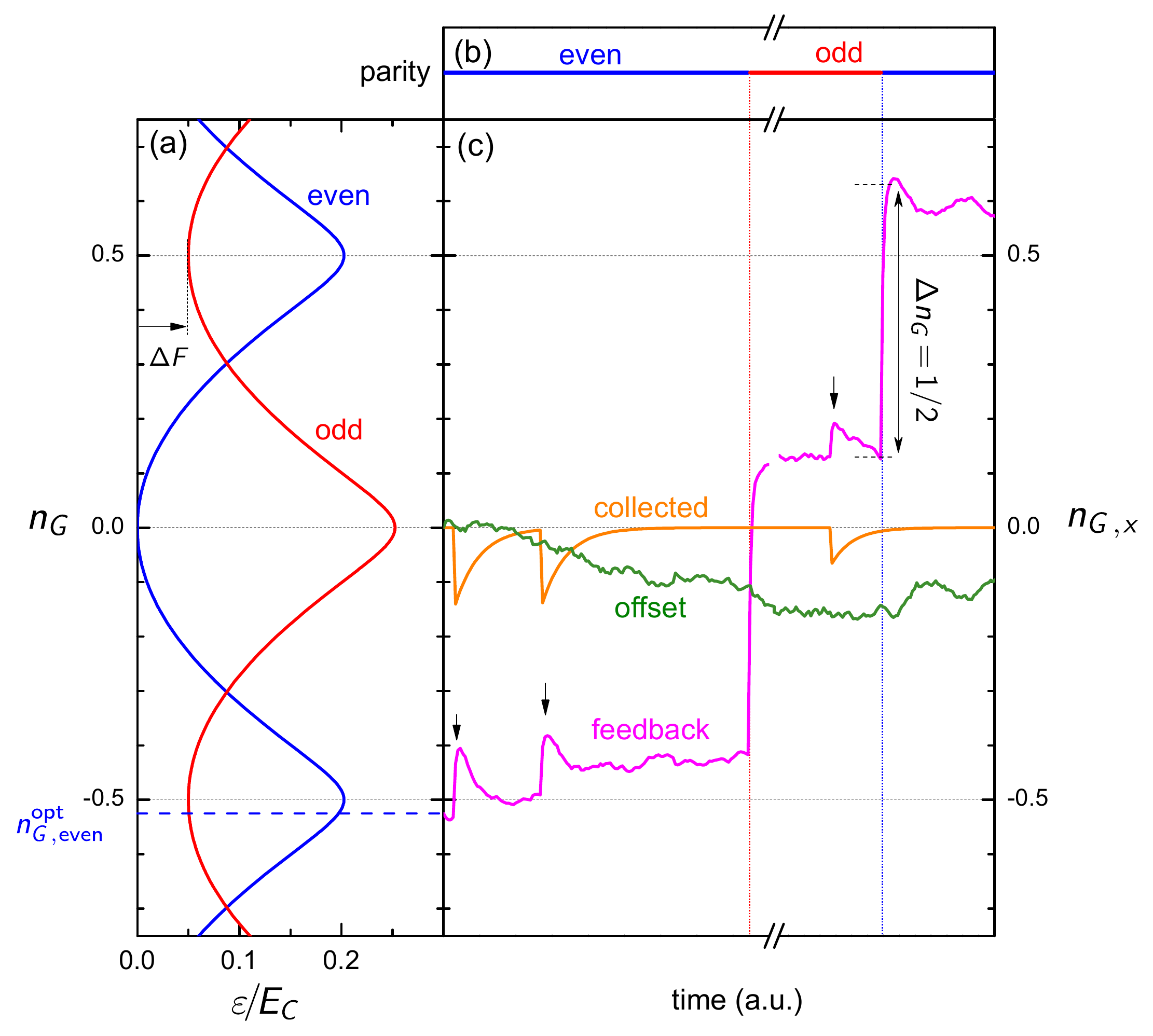}}
  \caption{\label{even-odd} Sketch of a measurement in
  feedback mode, in presence of parity jumps and $1 / f$ charge noise. (a)
  Even and odd ground state energy bands of the CPB as a function of total
  gate charge $n_G$ at $E_J / E_C = 0.1$ (swapped axes). (b,c) Sketch of a
  possible time variation of the island parity, offset charge drift, and
  collected charge on the detector gate. Corresponding feedback gate charge
  maintaining the amplitude and phase of the transmitted microwave signal,
  calculated with a finite feedback speed higher than the detector discharge
  speed. The three detected events are indicated by arrows. The detector is
  blind during the recovery time after every parity jump.}
\end{figure}

Although the volumic density $\rho_{\tmop{qp}}$ of quasiparticle excitations
in a superconductor should decrease with temperature $T$ as $\sim \exp (-
\Delta / k_B T)$ with $\Delta$ the superconducting energy gap, one always
observe at mK temperature out-of-equilibrium quasiparticles at a much higher
density, of the order of a few ones per $\mu m^3
${\cite{bespalov_theoretical_2016}}. These quasiparticles can tunnel across
the junction and poison the island with an unpaired electron. As a
consequence, the simple energy diagram of Fig. \ref{NRJ_Cq_of_nG}b valid for
the so-called ``even'' states corresponding to all electrons paired, becomes
the diagram of Fig. \ref{even-odd}a with an additional ``odd'' ground state
band shifted by $1 e$ in $n_G$ and by the free-energy difference $\Delta F$
{\cite{lafarge_measurement_1993}}. Whenever a quasiparticle tunnels, the
state of the island switches between the two bands of different parities
{\cite{lafarge_measurement_1993}}, which switches on and off the quantum
capacitance at $n^{\tmop{opt}}_G \simeq 1 / 2$. With the corresponding
switching rates $\Gamma_{o \rightarrow e}$ and $\Gamma_{e \rightarrow o}$ ,
the CPB spends a fraction of the time $p_e = \Gamma_{o \rightarrow e} /
(\Gamma_{o \rightarrow e} + \Gamma_{e \rightarrow o})$ in the good (i.e.
charge sensitive) even state.

The thermodynamics and the kinetics
{\cite{naaman_time-domain_2006,lutchyn_kinetics_2007,shaw_kinetics_2008-1}} of
these jumps are governed in particular by $\rho_{\tmop{qp}}$, and by the
density $\rho_{\tmop{sg}}$ of island subgap states. Using an island material
so that $\rho_{\tmop{sg}} $ is small, one can make $\Gamma_{e \rightarrow o}$
very slow at low temperature, as demonstrated with NbTiN islands and Al
grounds {\cite{van_woerkom_one_2015}}, for which about 1 minute even state
lifetime was observed. With such a lifetime much longer than the inverse
bandwidth of the electrometer discussed in section \ref{dS21dfNL}, and in
absence of specific feedback, the electrometer is operational only a fraction
$p_e$ of the time, as illustrated in Fig. \ref{even-odd}b. But it is also
possible to program a feedback that shifts $n^{}_G$ by $1 e$ whenever a
poisoning event occurs, in order to actively restore the even state
sensitivity.

\section{Integrating our electrometer in a macroscopic detector}

The range of applications of the electrometer proposed here is wide: it could
be used to study new quantum electrical devices or address present-day
questions in mesoscopic physics, like measuring the charge of Majorana split
states. However our goal is to use it for detection of rare astroparticle
events creating a bunch of charges inside a massive semiconducting cristal
(e.g Germanium) placed at a few mK.

The main problem in this bolometric application is the small coupling between
the massive block and the mesoscopic electrometer: indeed, the block having a
large self-capacitance compared to the CPB gate, only a small fraction ($\sim 10^{- 5}$) of the total created charge is collected onto the CPB gate. To
maximize the coupling, the detector gate capacitance between the CPB island
and the semiconducting block should be made large, although we have concluded
from our optimization that $E_C$ should be maximized. This means that the
total capacitance budget is tight, and that a maximum amount of this budget should be devoted to coupling to the semiconducting block.

A simple evaluation of this budget goes as follows: The total capacitance is
distributed between a Josephson junction of 100x100nm ($C_J \sim 500
\tmop{aF}$), a detector gate of similar capacitance ($C_{G, d} \sim 500
\tmop{aF}$), and the other gates and parasitic capacitances ($C_{G, r} + C_{G,
f} \sim 200 \tmop{aF}$), which results in a charging energy $E_C / k_B = 3 K$.
Taking a low-loss ($r = 0.1$) resonator of frequency $f_r = 2.4 \tmop{GHz}$,
we find an optimum Josephson coupling $E_J / E_C = 0.1$ and the corresponding
quantum limited sensitivity $S^B_e = 9.10^{- 7} e^- / \sqrt{\tmop{Hz}}$, with
a bandwidth determined by $\omega_0 / Q_t$.

A Ge bolometer detector such as the ones used in CDMS
{\cite{supercdms_collaboration_new_2016}} or EDELWEISS
{\cite{armengaud_constraints_2016}} can be optimized to have a capacitance
between electrodes of about $C_{\det} = 10 \tmop{pF}$. Taking the capacitive
division between $C_{\det}$ and $C_{G, d}$ into account yields an energy
sensitivity of $1.8 \tmop{eV} / \sqrt{\tmop{Hz}}$ for neutral events, assuming
an energy to charge conversion of $10 \tmop{eV} / e^-$, typical for nuclear
recoil in Germanium {\cite{main_note_2}}. This result outperforms by far (one
to two orders of magnitude for similar $C_{\det}$) all existing semiconducting
charge amplifiers, even when accounting for a larger capacitive division
{\cite{phipps_hemt-based_2016}}.

Now in terms of bare charge sensitivity of the electrometer alone, these
$9.10^{- 7} e^- / \sqrt{\tmop{Hz}}$ could seem disapointing compared to the
$\sim 10^{- 6} e^- / \sqrt{\tmop{Hz}}$ sensitivity reported for the best
superconducting $\mu$w-CPB or CPTs {\cite{brenning_ultrasensitive_2006}}.
However, we stress that ours is given for a geometry suitable for coupling to
macroscopic detectors, i.e. with a large coupling capacitance. Should we take
the island geometrical parameters of ref.
{\cite{brenning_ultrasensitive_2006}}, i.e. an island capacitance $58
\tmop{aF}$ ($E_C / k_B = 64 K$), we would predict a sensitivity one order of
magnitude better ($S^B_e = 0.3 \times \sqrt{(1 + r) \hbar / 2 E_C} \simeq 0.8
\cdot 10^{- 7} e^- / \sqrt{\tmop{Hz}}$), using the previous resonator
parameters and the new optimum $E_J$ (along the hyperbola $\hbar \omega_r /
E_C = 0.002$).

\section{Conclusions}

We have shown how to optimize a wide-band $\mu$w-Cooper Pair Box
electrometer. By taking advantage of the resonator nonlinearity induced by the
CPB, a quantum limited sensitivity about ten times better than the linear one
can be obtained, similar to what would be achieved using a separate quantum
limited amplifier.

Interestingly, in the nonlinear case, the CPB-resonator coupling strength and
the resonator impedance disappear from the optimization problem, which leads
to a much simpler constrained numerical optimization.

Taking routinely achiveable parameters, we predict a conservative quantum
limited sensitivity $S^B_e$ in the $10^{- 7} e^- / \sqrt{\tmop{Hz}}$ range,
which should make possible to clarify the physics of new quantum electrical
devices, and to fabricate much more sensitive bolometers for particle
detection.

\section{Acknowledgements}

The authors are warmly thankful to D. Est{\`e}ve, P. Joyez and the Quantronics
group for stimulating discussions about the CPB, and thank L. Dumoulin and S.
Marnieros for sharing their expertise on bolometric dark matter detection. HLS
aknowledges support from the FCS Campus Paris-Saclay under the project
2011-019T-COCA, from the ANR JCJC grant ANR-12-JS04-0007-01, and from the
PNCG.

\end{document}